\documentclass[prc,aps]{revtex4}
\usepackage{graphicx}
\begin{document}

\title{A new puzzle for random interaction}

\author{Calvin W. Johnson and Hai Ah Nam}
\affiliation{Physics Department, San Diego State University, 5500
Campanile Drive, San Diego, California 92182-1233}

\begin{abstract}
We continue a series of numerical experiments on many-body systems
with random two-body interactions, by examining correlations in
ratios in excitation energies of yrast $J$ = 0, 2, 4, 6, 8
states. Previous studies, limited only to $J$ = 0,2,4 states, had
shown strong correlations in boson systems but not fermion systems.
By including $J \ge 6$ states and considering different scatter
plots, strong and realistic correlations appear in both boson and
fermion systems. Such correlations are a challenge to 
explanations of random interactions.
\end{abstract}

\maketitle

Nuclei show a remarkable array of behaviors at low
excitation energy, notably collective motion for
even-even nuclides\cite{BM}.  Three sets of tools have emerged
to shed light on collective motion: algebraic models,
based on representations of low-dimension groups in
many-fermion and many-boson systems\cite{simple};
precise characterizations of the nucleon-nucleon interaction
and rigorous derivation of effective interactions\cite{abinitio}; and,
paradoxically, studies of the behavior of random interactions\cite{JBD98}.

The study of random two-body interactions was originally applied to
quantum chaos and statistical properties of compound nuclear states\cite{FW70}.
A few years ago numerical experiments showed, surprisingly, that
random interactions could show spectral signatures of regular,
collective behavior\cite{JBD98}. The first and foremost signature is that, out
of an ensemble of randomly chosen two-body interactions, the ground
state predominantly has angular momentum $J=0$ (typically 50-70$\%$)
even though such states are a small fraction (typically 2-10$\%$) of
the many-body space. There are other signatures, which we will not
review fully here.

Instead we focus on band structure, in particular vibrational and
rotational bands, experimentally seen in many even-even nuclides and
which in part lead to the liquid drop model and its quantized
version the collective geometric model (a generalization of the Bohr
Hamiltonian).  The most obvious signature of bands are regular
structures in the excitation energies of yrast $J$= 0, 2, 4 ..
states: archetypal vibrational bands have excitation energy $E_x(J)
\propto J$ while rotational bands have $E_x(J) \propto J(J+1)$. (A
deeper, and no less important signature, are ratios of intraband E2
transitions strengths.) The simplest model of pairing, the seniority model, by 
contrast has the first excited $J=0,2,4,...$ states degenerate.

Bijker and Frank\cite{BF00} found strong evidence for band structure in the
interacting boson model (IBM) with random interactions through two
pieces of evidence. First, they profiled, for an ensemble of random
interactions, the frequency of the excitation energy ratio
\begin{equation}
R_{42} \equiv E_x(J=4)/E_x(J=2) \label{r42}
\end{equation}
and found sharp peaks at $R_{42}$ = 2 and 3.33, corresponding to
vibrational and rotational bands. (The seniority model has 
$R_{42} = 1 $.)  More significantly, they made a
scatter plot of $R_{42}$ versus the ratio of E2 transition
strengths $B(E2:4_1^+ \rightarrow 2_1^+)/B(E2:2_1^+ \rightarrow 0_1^+)$, 
and found significant enhancements at the exact U(5)
(vibrational) and SU(3) (rotational) limits.  Bijker and Frank later
analyzed these results in terms of a mean-field model\cite{BF01}.

Fermion systems with random interactions do not show the same
correlations as boson systems. Fig.~1 shows frequency
distributions for $R_{42}$ for several typical cases. Fig.~1(a) 
is for 10 identical particles in a $1p_{1/2}$-$1p_{3/2}$-$0f_{5/2}$-$0f_{7/2}$ 
or $pf$ space, which, if one assumed a closed $^{40}$Ca core, would
correspond to $^{50}$Ca; of course, the interaction is random and 
there is no \textit{a priori} constraint on the single-particle radial 
wavefunctions, so labelling this system as $^{50}$Ca is simply for 
convenience. In this and all cases the interaction conserves 
angular momentum. Fig.~1(b) has 4 protons ($\pi$) and 4 neutrons ($\nu$) 
in a $1s_{1/2}$-$0d_{3/2}$-$0d_{5/2}$ or $sd$ space, so we 
colloquially refer to it as $^{24}$Mg; in this case we constrain the 
interaction to conserve isospin as well. For the final two panels 
of Fig.~1 we use single-$j$ spaces, popular with many investigations 
of random interactions. Fig.~1(c) has 10 identical particles in 
a $j = 21/2$ space while 1(d) has 4 protons and 4 neutrons in a 
$j=13/2$ space. These systems were chosen in order to have a large number 
of interacting particles and non-trivial dimensions of the many-body space 
($M$-scheme dimension of about $10^4$), but still relatively small 
enough that a high-performance $M$-scheme shell model code \cite{redstick} 
can run thousands of cases. For all the plots in this paper we only select 
cases for which the ground state has $J=0$.

 The broad
peaks at $R_{42} =1$ in Fig.~1 is closest to a simple seniority model, and
although other signatures of pairing can be found \cite{JBDT00}, detailed investigations discourage
interpretation as a simple pairing condensate \cite{ZV04}. 
Similarly, a Bijker-Frank plot of $R_{42}$ versus ratio of
E2 strengths shows no strong correlations (not shown here).

\begin{figure}
\includegraphics[scale=0.43]{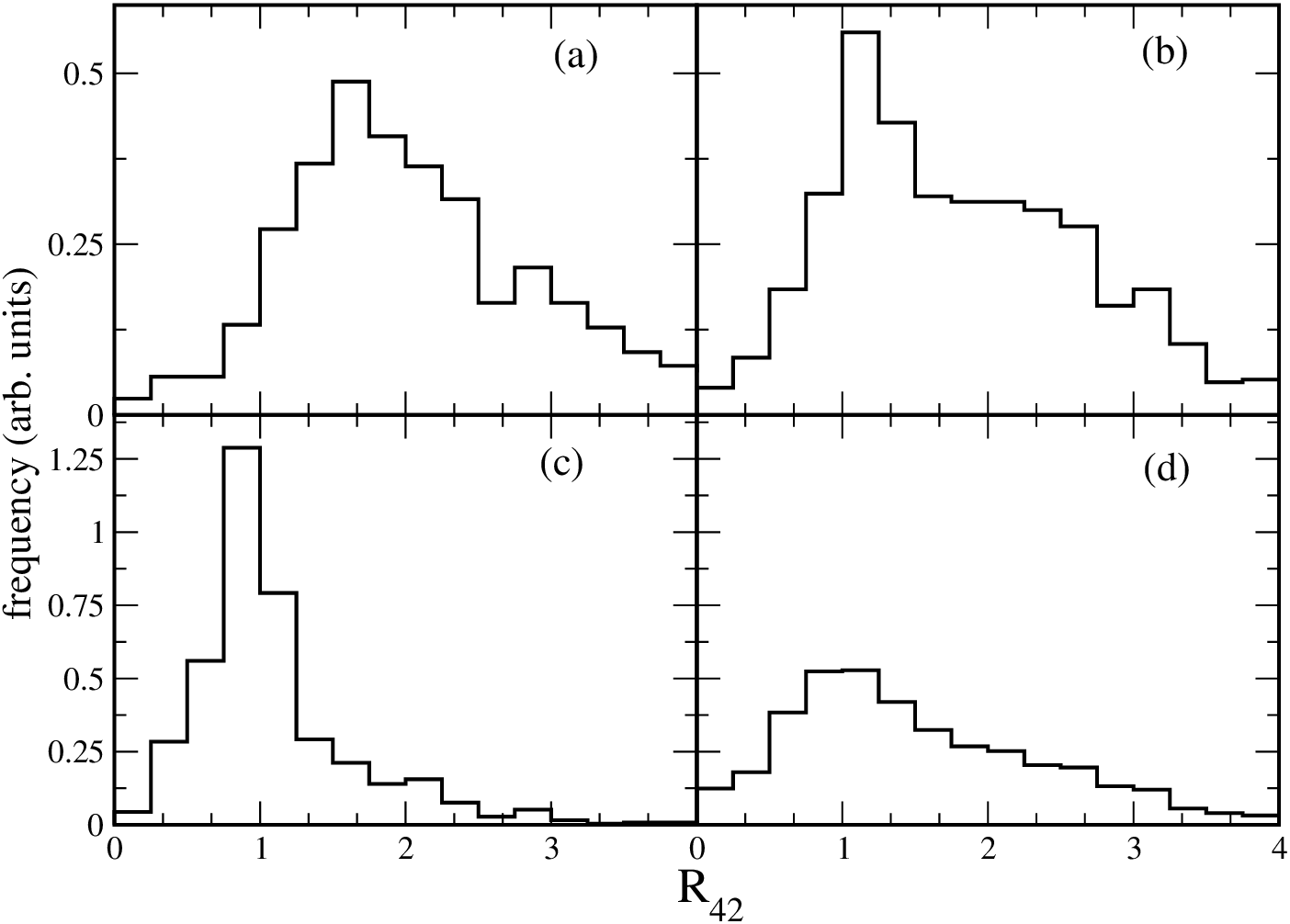}
\caption{\label{r42freq} Frequency distribution of 
$R_{42}$ for (a) $(pf)^{10}$ (``$^{50}$Ca''); (b) $(sd)_\pi^4 (sd)_\nu^4$
(``$^{24}$Mg''); (c) $(21/2)^{10}$; and (d) $(13/2)_\pi^4 (13/2)_\nu^4$
}
\end{figure}

Although random interactions do not  quantitatively reproduce
experimental behavior, the results are striking enough to have
spawned a mini-industry. Several attempts have been made to explain the results; 
see \cite{ZAY04,ZV04,PW04} for some broad examples. Although these analyses yield 
some valuable and interesting insights, arguably none rise to the level 
of a comprehensive ``theory'' of random interactions complete with 
predicting new phenomena. Thus there is still room for empirical 
exploration.

In this short note, we show results of a new numerical 
experiment.  In particular, we look at correlations with 
higher-$J$ yrast states, which with few exceptions \cite{ZV04} 
have been paid scant attention so far.

We define $R_{62} \equiv E_x(J=6)/E_x(J=2)$, as well as 
$R_{82}$, etc. in an obvious generalization to Eq.~(\ref{r42}); in Figs. 2 and 3 we 
look at scatter plots of $R_{62}$ vs. $R_{42}$, and $R_{82}$ vs. $R_{42}$, 
respectively. We also show the loci for seniority, vibrational, and 
rotational bands. Given the broad structure in Fig. 1, these strong correlations
are surprising.

\begin{figure}
\includegraphics[scale=0.43,clip]{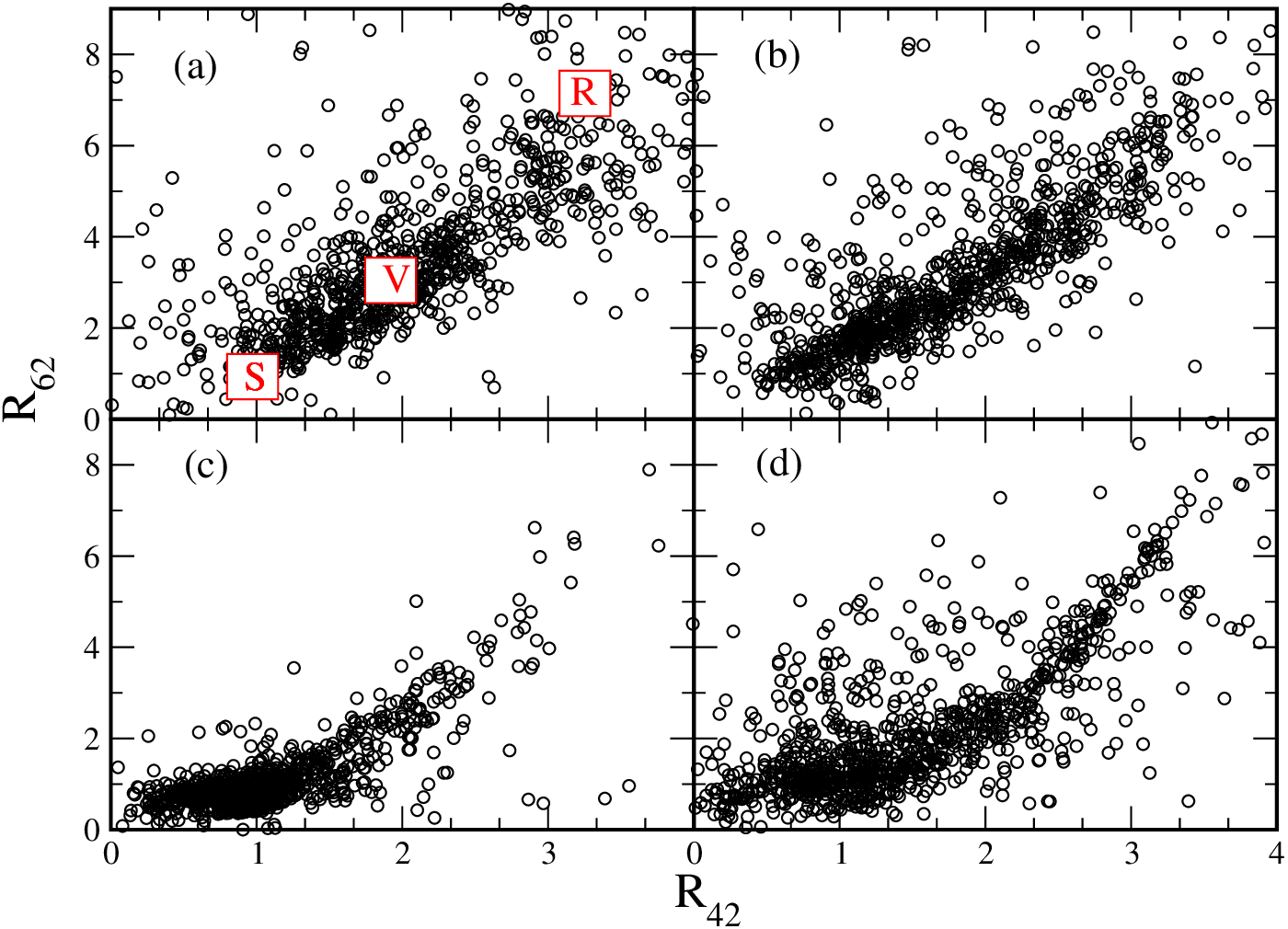}
\caption{\label{r42_r62} (Color online) Scatter plot of 
$R_{62}$ versus $R_{42}$ for (a) $(pf)^{10}$ (``$^{50}$Ca''); (b) $(sd)_\pi^4 (sd)_\nu^4$
(``$^{24}$Mg''); (c) $(21/2)^{10}$; and (d) $(13/2)_\pi^4 (13/2)_\nu^4$. 
Also shown are locations for seniority (`S'), vibrational (`V'), and rotational (`R') 
limits.
}
\end{figure}

Although there are strong correlations for the energy spectrum in fermion 
systems, another signal of band structure are large, consistent quadrupole deformations.
We looked for these by computing the quadrupole moments of the yrast $J > 0 $ states and 
found no obvious correlations. We also looked at ratios of B(E2) strengths 
\cite{jn06} and found only weak correlations.

\begin{figure}
\includegraphics[scale=0.43,clip]{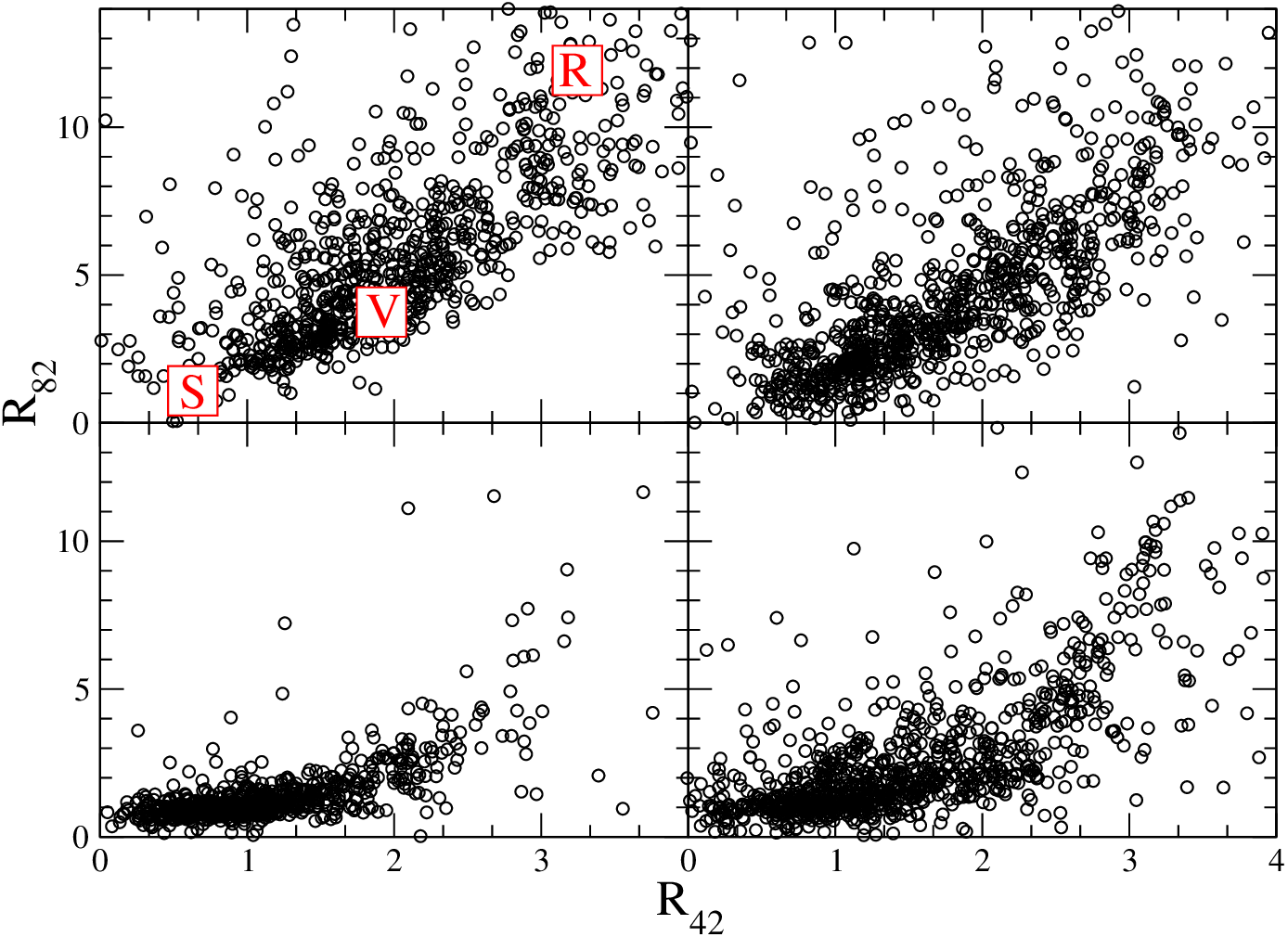}
\caption{\label{r42_r82} (Color online) Scatter plot of 
$R_{82}$ versus $R_{42}$ for (a) $(pf)^{10}$ (``$^{50}$Ca''); (b) $(sd)_\pi^4 (sd)_\nu^4$
(``$^{24}$Mg''); (c) $(21/2)^{10}$; and (d) $(13/2)_\pi^4 (13/2)_\nu^4$. 
Also shown are locations for seniority (`S'), vibrational (`V'), and rotational (`R') 
limits.
}
\end{figure}

Finally in Fig.~4 we show similar results for the interacting boson model
with random interaction, using the program PHINT\cite{phint}. 
Similar strong correlations, not shown, occur even for $J = 10$. 
Here are the correlations are less surprising, given the results of Bijker 
and Frank. 

\begin{figure}
\includegraphics[scale=0.40,clip]{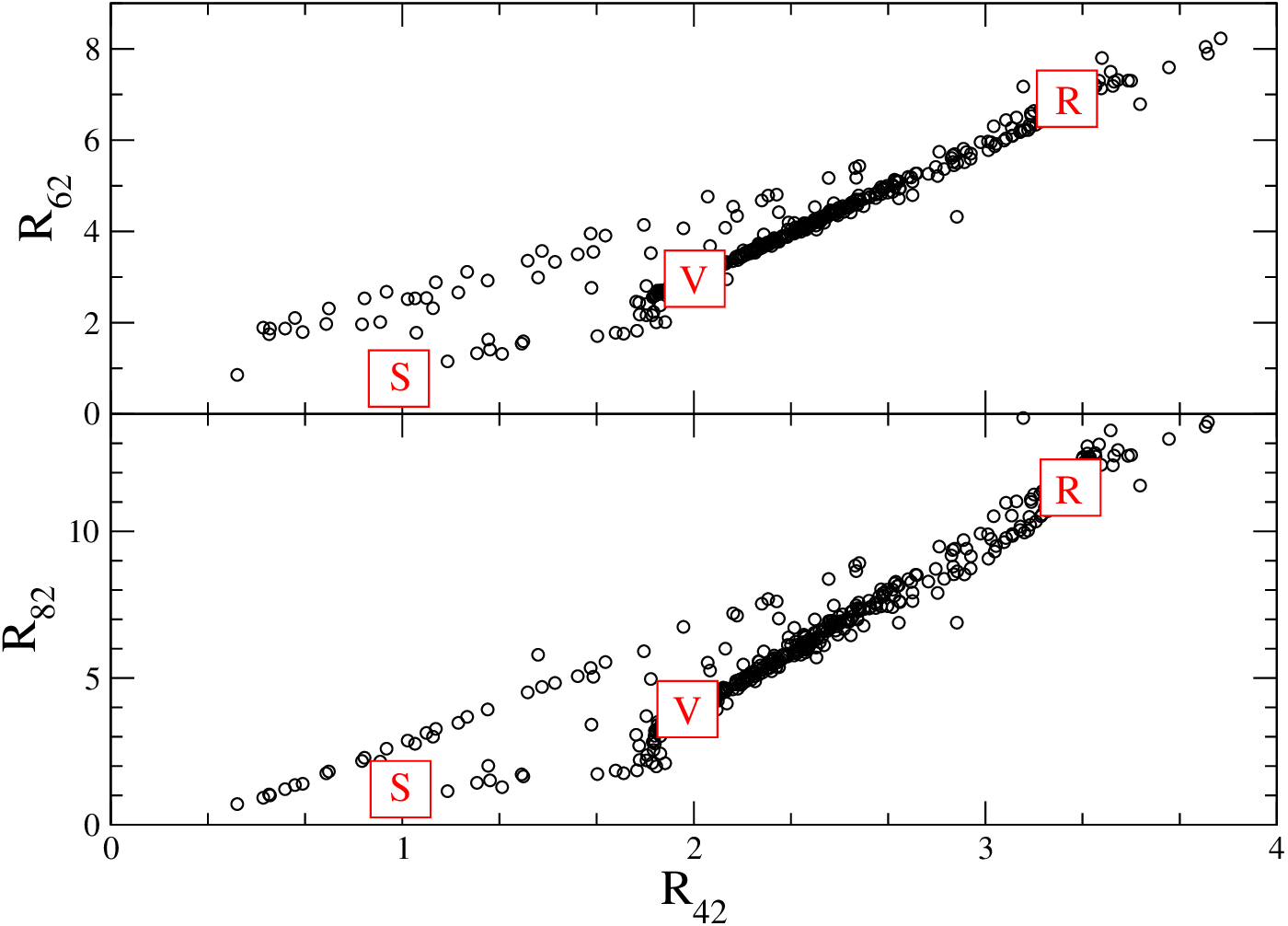}
\caption{\label{boson} (Color online) Scatter plot of 
(a) $R_{62}$ versus $R_{42}$ and (b) $R_{82}$ versus $R_{42}$ for
the interacting boson model with 16 bosons. Also shown are locations for 
seniority (`S'), vibrational (`V'), and rotational (`R') limits.
}
\end{figure}

 We have no broad explanation for these results, and certainly no 
quantitative explanation. One can invoke a mean-field picture, but given 
the presence of ``geometric chaoticity'' \cite{ZV04} it is surprising that 
only a select range of mean-fields can form. Other recent work \cite{PW04} has 
shown that shell-model dynamics are dominated by a relatively few combinations 
of two-body matrix elements; why such select combinations give rise to 
collectivity or even pseudo-collectivity is not immediately clear and not 
addressed by those authors. For the moment we present these curious 
\textit{empirical} phenomena as a provocative challenge to existing 
and future analyses of random interactions.

The U.S.~Department of Energy supported this investigation through
grant DE-FG02-96ER40985.

\end{document}